\documentclass[twocolumn,dvipsnames]{aastex62}
\usepackage{CJK}
\usepackage{framed}

\def\lsim{\lesssim}
\def\Msun{{\rm M}_\odot}
\def\Muv{$M_{1500}$}
\def\dim#1{{\rm\,#1}}

\graphicspath{{./}{figures/}}

\received{December 2, 2019}
\accepted{July 1, 2020}
\submitjournal{ApJ}

\shorttitle{CROC Galaxy-Halo Connection}
\shortauthors{Zhu et al.}

\begin{document}
\begin{CJK*}{UTF8}{gkai}

\title{Cosmic Reionization On Computers: The Galaxy-Halo Connection between $5 \leq z \leq10$}

\correspondingauthor{Hanjue Zhu (朱涵珏)}
\email{hanjuezhu@uchicago.edu}
\author[0000-0003-0861-0922]{Hanjue Zhu (朱涵珏)}
\affiliation{The University of Chicago;
Chicago, IL 60637, USA}

\author[0000-0001-8868-0810]{Camille Avestruz}
\affiliation{Leinweber Center for Theoretical Physics; Department of Physics; University of Michigan, Ann Arbor, MI 48109, USA}

\author{Nickolay Y. Gnedin}
\affiliation{Particle Astrophysics Center; 
Fermi National Accelerator Laboratory;
Batavia, IL 60510, USA}
\affiliation{Kavli Institute for Cosmological Physics;
The University of Chicago;
Chicago, IL 60637, USA}
\affiliation{Department of Astronomy \& Astrophysics; 
The University of Chicago; 
Chicago, IL 60637, USA}

\begin{abstract}
We explore the connection between the stellar component of galaxies and their host halos during the epoch of reionization ($5 \leq z\leq10$) using the CROC (Cosmic Reionization on Computers) simulations. We compare simulated galaxies with observations and find that CROC underpredicts the abundance of luminous galaxies when compared to observed UV luminosity functions, and analogously the most massive galaxies when compared to observed stellar mass functions.  We can trace the deficit of star formation to high redshifts, where the slope of the star formation rate to stellar mass relation is consistent with observations, but the normalization is systematically low.  This results in a star formation rate density and stellar mass density that are systematically offset from observations.  However, the less luminous or lower stellar mass objects have luminosities and stellar masses that agree fairly well with observational data. We explore the stellar-to-halo mass ratio (SHMR), a key quantity that is difficult to measure at high redshifts and that models do not consistently predict.  In CROC, the SHMR {\it decreases} with redshift, a trend opposite to some abundance-matching studies. These discrepancies uncover where future effort should be focused in order to improve the fidelity of modeling cosmic reionization. We also compare the CROC galaxy bias with observational measurements using Lyman-break galaxy samples, finding reasonable consistency.
\end{abstract}


\section{Introduction}\label{sec:intro}

Observations of high-redshift galaxies inform modeling of galaxy formation during the epoch of reionization (e.g. \citealt{bradley12}, \citealt{schmidt14}, \citealt{zitrin14}, \citealt{zitrin15}, \citealt{oesch15}, \citealt{oesch16}, \citealt{bernard16}, \citealt{livermore17}, \citealt{ishigaki18}, \citealt{morishita18}).  Current observations now robustly constrain the faint-end slope of the UV luminosity functions at $z\gtrsim6$ . These observations come from gravitationally lensed images of the most distant galaxies, but lensing measurements have systematic uncertainties associated with them \citep{wyithe11}. Different measurements of the stellar mass function at these redshifts also have considerable discrepancies.  Uncertainties in stellar mass measurements come from both the intrinsic scatter in empirical relations and uncertainties in model-dependent assumptions used to derive the stellar mass.

Future observational facilities, such as the James Webb Space Telescope (JWST; \citealt{Gardner2006}) and next-generation ground-based telescopes, will increase the sample of high-redshift galaxies and push observational limits to fainter values. Data from these telescopes will help determine the faint-end slopes of the UVLF and the SMF. To maximize the impact of observations, we need robust theoretical models to better predict high-redshift galaxy properties and aid in the interpretation of upcoming data.

Galaxy formation is difficult to model because it incorporates nonlinear interplay between a wide range of physical processes, such as gas cooling, star formation, and stellar feedback processes. On the other hand, the growth of large-scale structure and the evolution of dark matter halos can be predicted robustly using numerical simulations in the $\Lambda$CDM model of cosmology. Dark matter halos grow through cosmic accretion and mergers with other halos. Galaxies form within dark matter halos, forming stars from gas reservoirs, and growing through both accretion of gas and stars and galaxy-galaxy mergers. Given the robust results from dark-matter-only simulations and the complexity of galaxy formation processes, it is natural for us to explore the connections between observed galaxies and the underlying dark matter distribution (see \citet{wechsler18} for a recent review). Galaxy-halo connections encapsulate the physical processes driving galaxy formation. This connection generally illustrates empirical correlations between galaxy properties and the properties of the extended dark matter halos that host galaxies (e.g. \citealt{conroy06}, \citealt{white07}, \citealt{zheng07}, \citealt{firmani10}, \citealt{tinker13}, \citealt{wang13}, \citealt{birrer14}, \citealt{sun16}, \citealt{cohn17}, \citealt{mitra17}). 

Recent observations have also measured clustering properties of high-redshift galaxies as a way to further connect observed galaxies with their host dark matter halos (\citealt{baronenugent_etal14}, \citealt{sobacchi15}, \citealt{harikane18}). The clustering of dark matter halos is easily obtained from numerical simulations.  However, galaxy clustering depends on the detailed relationship between the galaxies and their dark matter host halos.  Clustering statistics can therefore provide additional constraints to galaxy formation models.

There are two kinds of approaches to model the galaxy-halo connection: empirical modeling and physical modeling. Empirical modeling uses data to constrain parameters that describe the galaxy-halo connection; physical modeling either parameterizes (semi-analytical modeling) or directly simulates the physics of galaxy formation using cosmological hydrodynamical simulations.
Simulations solve for gravity and hydrodynamics, while incorporating extensive physical prescriptions for galaxy formation processes.

The Cosmic Reionization on Computers (CROC) project produces such cosmological simulations that self-consistently model relevant
physical processes in cosmological volumes during the epoch of reionization. CROC therefore connects the dynamics of dark matter and galaxy formation.

In this paper, we first look at how CROC outputs compare to the observed luminosity and stellar mass properties; we show the CROC UVLF in Section~\ref{sec:results:uvlf}, the stellar mass-to-light ratio in Section~\ref{sec:results:masstolight}, and the SMF in \ref{sec:results:stellarmassfunction}. We additionally quantify the galaxy-halo connection in CROC by looking at the stellar-to-halo mass ratio (SHMR) in \ref{sec:results:galaxytohalo}. Subsequently, we show the star formation histories in \ref{sec:results:starformationhistory}. Finally, we use the relationship between the UV luminosity and halo mass in CROC to calculate the galaxy bias, or the excess clustering of galaxies compared with that of the underlying matter, and compare with observational results from \citet{baronenugent_etal14} in Section~\ref{sec:results:bias}. The galaxy bias complements the other tests of the galaxy-halo connection.

\section{Methodology}\label{sec:methods}

\subsection{CROC Simulations}

All CROC simulations were performed with the Adaptive Refinement Tree (ART) code \citep{kravtsov99,kravtsov_etal02,rudd_etal08}. They include a wide range of physical processes that are thought to be necessary for self-consistent modeling of cosmic reionization, such as gravity and gas dynamics, fully coupled radiative transfer, atomic cooling and heating processes, molecular hydrogen formation, star formation and stellar feedback. Full details of the simulations are described in the  CROC method paper \citep{gnedin14}. For stellar population syntesis modeling, we use the Flexible Spectral Population Synthesis code \citep[FSPS;][]{fsps}.

In this paper we use two sets of simulations: a set of six independent random realizations in $L_{box}=40h^{-1} \dim{Mpc}$ comoving boxes with $1024^3$ dark matter particles and a set of three independent random realizations in $L_{box}=80h^{-1} \dim{Mpc}$ comoving boxes with $2048^3$ dark matter particles. Both simulation sets have the same spatial resolution of $100 \dim{pc}$ in proper units (kept constant throughout the simulation) and effective mass resolution of $M_1=7\times10^6\Msun$ (defined as the mass of a dark matter particle in an equivalent dark-matter-only simulation).

\subsection{Halo Finder}

We use the ``yt" package to identify dark matter halos in the simulations \citep{yt_11}; yt supports two different halo finders: HOP \citep{EisensteinandHut_97} and Rockstar \citep{rockstar}. We have compared results using the  $L_{box}=40h^{-1} \dim{Mpc}$ boxes, which have halo catalogs from both halo finders. Both halo finders produce virtually indistinguishable results for all relations discussed in subsequent sections. HOP catalogs are available for all CROC simulation boxes; Rockstar catalogs are available for all $L_{box}=40h^{-1} \dim{Mpc}$ boxes but not for $L_{box}=80h^{-1} \dim{Mpc}$ due to numerical limitations of the Rockstar implementation on Blue Waters supercomputer. Unless specifically mentioned, by default we use HOP catalogs for the $L_{box}=80h^{-1} \dim{Mpc}$ dataset. 

Note that the halo mass output by HOP we choose is the mass within a radius enclosing an average density that is 160 times the critical density of the universe. The halo mass selected from the Rockstar catalogs uses the mass definition from \citet{Bryan98}: the mass within a radius enclosing an average density 360 times the background density at $z = 0$, including unbound particles.

\section{Results}\label{sec:results}

\subsection{The UVLF}\label{sec:results:uvlf}

We first compare the CROC galaxy UVLF with observational measurements in the literature. Since the UVLF is relatively well measured all the way to $z\sim10$, it is a natural choice for comparing theoretical predictions with observations. Galaxy luminosities at $1500$\AA\ (denoted as \Muv) are computed using the FSPS code \citep{fsps1,fsps2,fsps3}, including an observationally calibrated dust correction, as described in the first CROC paper (Appendix B of \citealt{gnedin14}). 

\begin{figure}
    \centering
        \includegraphics[width=0.99\columnwidth]{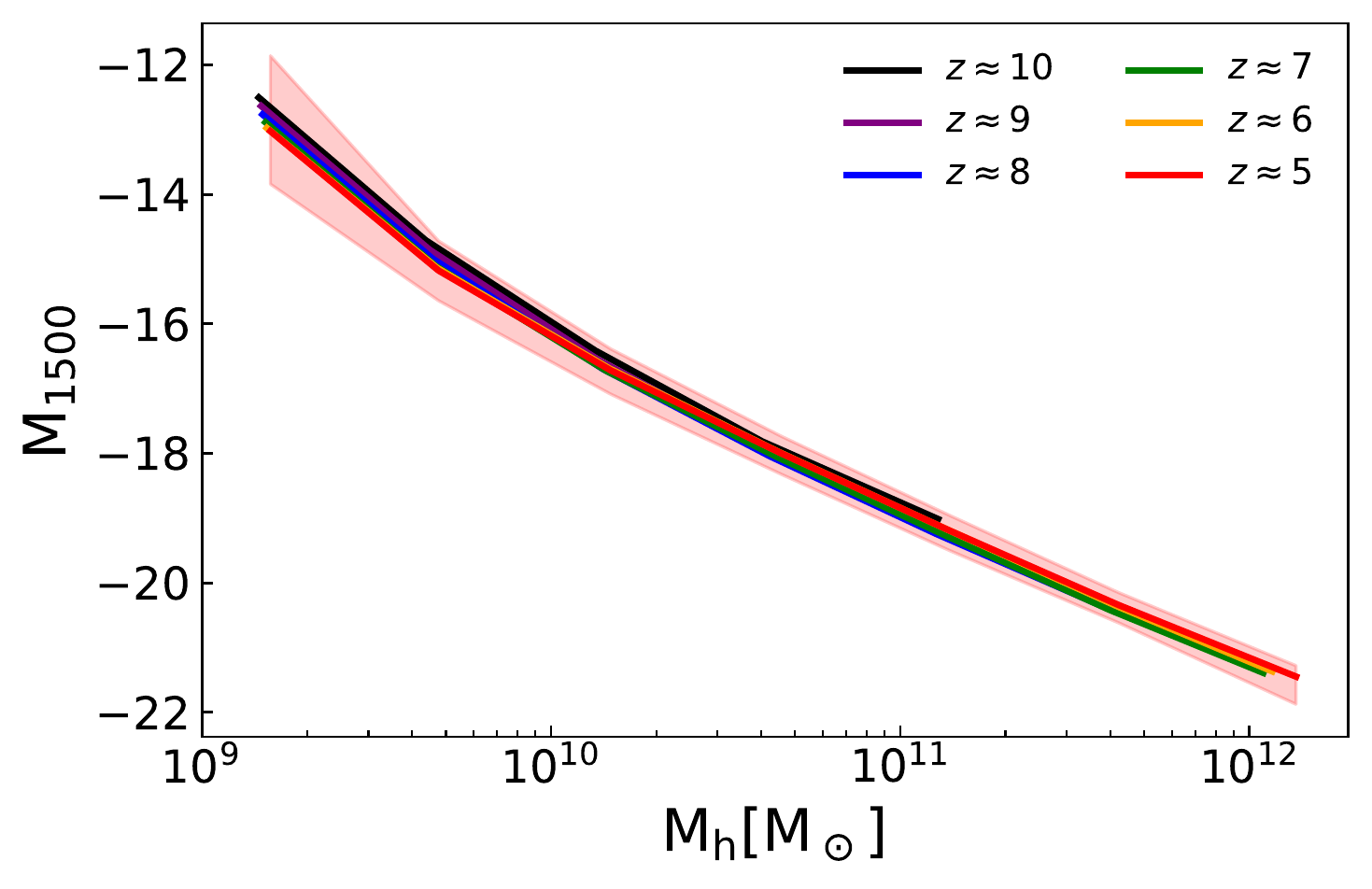} 
\caption{\Muv-$M_h$ relation in CROC from $z=5$ to $z=10$. The red transparent band illustrates the 25th and 75th percentile values in each halo mass bin at $z=5$ and is typical for the scatter in the relation at other redshifts as well.} \label{fig:muv_mh}
\end{figure}

The UVLF is shaped by the relation between the UV luminosity of simulated galaxies and the host halo mass. This relation, shown in Figure~\ref{fig:muv_mh} for $5\leq z\leq 10$, is rather tight, with the scatter being under 0.7 dex for most of the sampled range, and only increasing for the lowest-mass galaxies. An interesting feature of Figure~\ref{fig:muv_mh} is that the \Muv-$M_h$ relationship in CROC exhibits almost no redshift dependence. The star formation and stellar feedback model used in the simulations does not include any explicit assumption that can produce such an effect, so the lack of redshift dependence is a genuine predictions of the simulations. The significance of this prediction is, however, unclear at present.

\begin{figure}
    \centering
        \includegraphics[width=0.99\columnwidth]{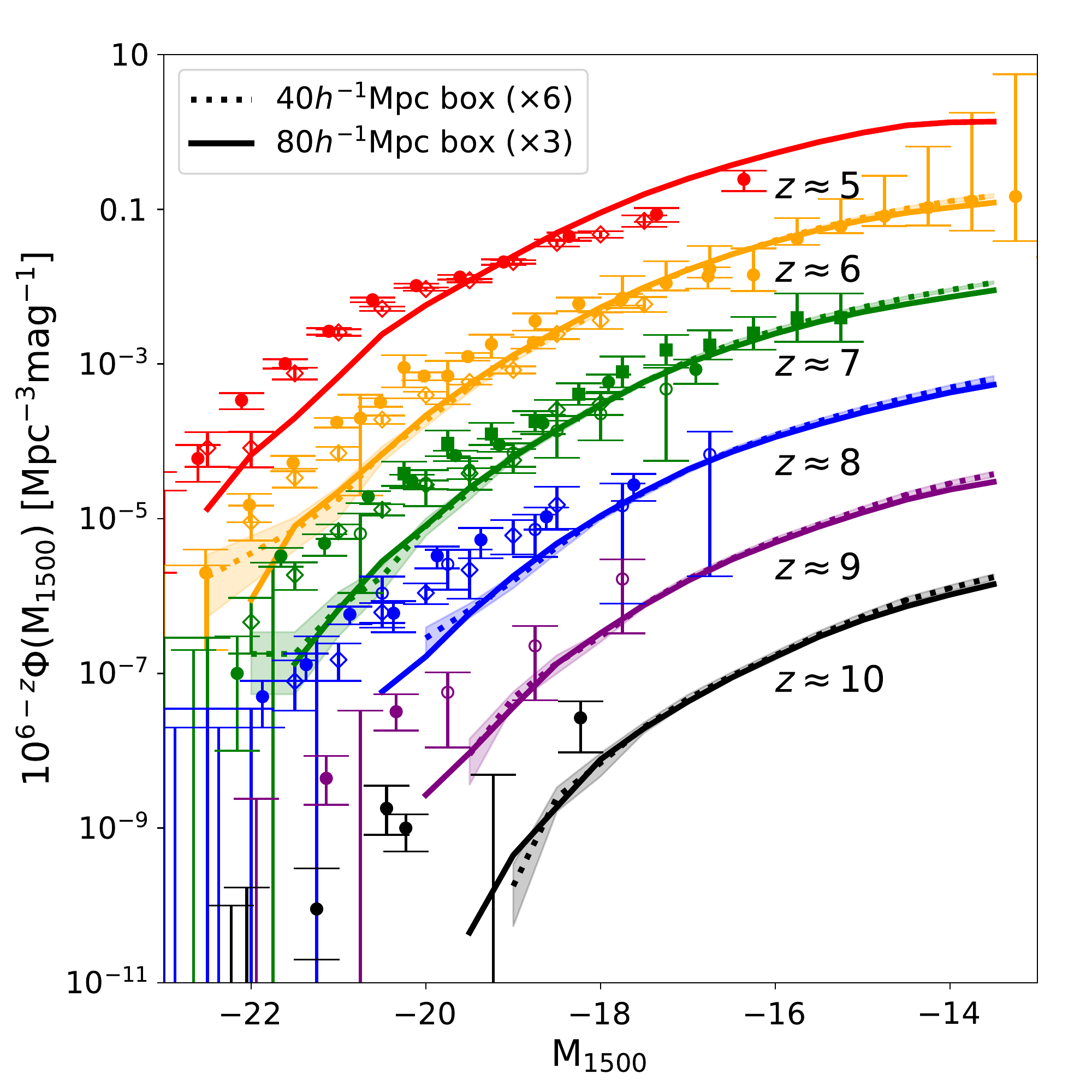} 
\caption{Comparison between the UV luminosity function (UVLF) from the CROC simulation suite boxes and observations. We manually apply an offset of 1 dex to each redshift to facilitate comparison between CROC and observational data at each redshift. The dotted lines correspond to the UVLF from the six 40$h^{-1} \dim{Mpc}$ comoving boxes in our suite.  Solid lines correspond to the UVLF from the three 80$h^{-1} \dim{Mpc}$ comoving boxes. The transparent bands illustrate the 25th and 75th percentile values in each UV luminosity bin for the 40$h^{-1} \dim{Mpc}$ boxes. Such bands for 80$h^{-1} \dim{Mpc}$ boxes are thinner than the width of the median line. The two sets of boxes agree well at all redshifts shown.  The observational data are from \citet{atek_etal15} (filled squares), \citet{bouwens_etal15, bouwens_etal17} (filled circles), \citet{finkelstein_etalb15} (open diamonds), and \citet{laporte_etal16} (open circles).  While there is general agreement between the CROC predictions and observations, CROC systematically underpredicts current measurements at magnitudes $M_{1500}\lsim -19$.}\label{fig:uvlf}
\end{figure}

Figure~\ref{fig:uvlf} shows the UVLF between $5\leq z\leq 10$ for CROC galaxies from all $L_{box}=80h^{-1} \dim{Mpc}$ and $L_{box}=40h^{-1} \dim{Mpc}$ boxes. We overplot a compilation of observed values. There is general agreement within the observational measurements at low and moderate luminosities, but CROC simulations systematically underpredict luminosities of brighter galaxies with magnitudes \Muv$\lsim -19$. Since the results from the 80$h^{-1} \dim{Mpc}$ comoving and 40$h^{-1} \dim{Mpc}$ comoving boxes agree fairly well, the source of the discrepancy is not in the limited volume of the CROC simulations.

The primary reason for the discrepancy, therefore, is the model for star formation and stellar feedback used in CROC. As discussed in \citet{gnedin14}, this model assumes a linear Kennicutt-Schmidt relation.
\begin{equation}
  \dot{\Sigma}_* = \frac{\Sigma_{\rm H_2}}{\tau_{\rm DEP}},
\end{equation}
where the molecular gas depletion time $\tau_{\rm DEP}=1.5\dim{Gyr}$ is taken as a universal constant, and $\Sigma_{\rm H_2}$ is computed using the fitting formulae from \citet{gd14}. Stellar feedback is modeled with a ``blastwave" approximation \citep{bw1,bw5,bw2,bw3,bw6,bw4}.

The physical effects that are not modeled in CROC simulations, such as active galactic nucleus (AGN) feedback and cosmic ray feedback, are expected to reduce the abundance of luminous galaxies \citep{sijacki15,kaviraj17}. The inclusion of these additional processes would likely exacerbate, rather than resolve, the discrepancy with observations. The exact cause of the suppression of star formation in massive galaxies in CROC simulations is unclear. Such a cause may lie in the non-constancy of the molecular gas depletion time; this is less likely as the depletion time is a proxy to star formation efficiency, and it is well established that self-regulation makes star formation rates in the simulations independent of the assumed value of efficiency \citep{Hopkins2018,Orr2018,Semenov2018}. For the same reason the detailed model for molecular hydrogen production (which serves as another proxy for star formation efficiency) should not matter for massive galaxies that are expected to be in the self-regulation regime. Another possibility is the excessive strength of stellar feedback in the CROC model. The assumed feedback strength does indeed affect the overall star formation rates in simulated galaxies \citep{Orr2018,Semenov2018} approximately proportionately; however, stronger feedback usually affects less massive galaxies more, while in our case it is in the most massive galaxies that the star formation rate is mis-modeled most significantly. These considerations serve as a motivation for further testing of the sub-grid models in CROC.

\subsection{The Stellar Mass-to-light Ratio}\label{sec:results:masstolight}

In this section, we examine the stellar mass-to-luminosity ratio, or the stellar mass-to-light ratio, as a function of galaxy stellar mass.  We plot this instead of the $M_{\star}-L_{\star}$ relation to reduce the dynamic range and better illustrate modest trends. Note that the two relations are interchangeable.  

\begin{figure}[t]
\centering
        \includegraphics[width=0.99\columnwidth]{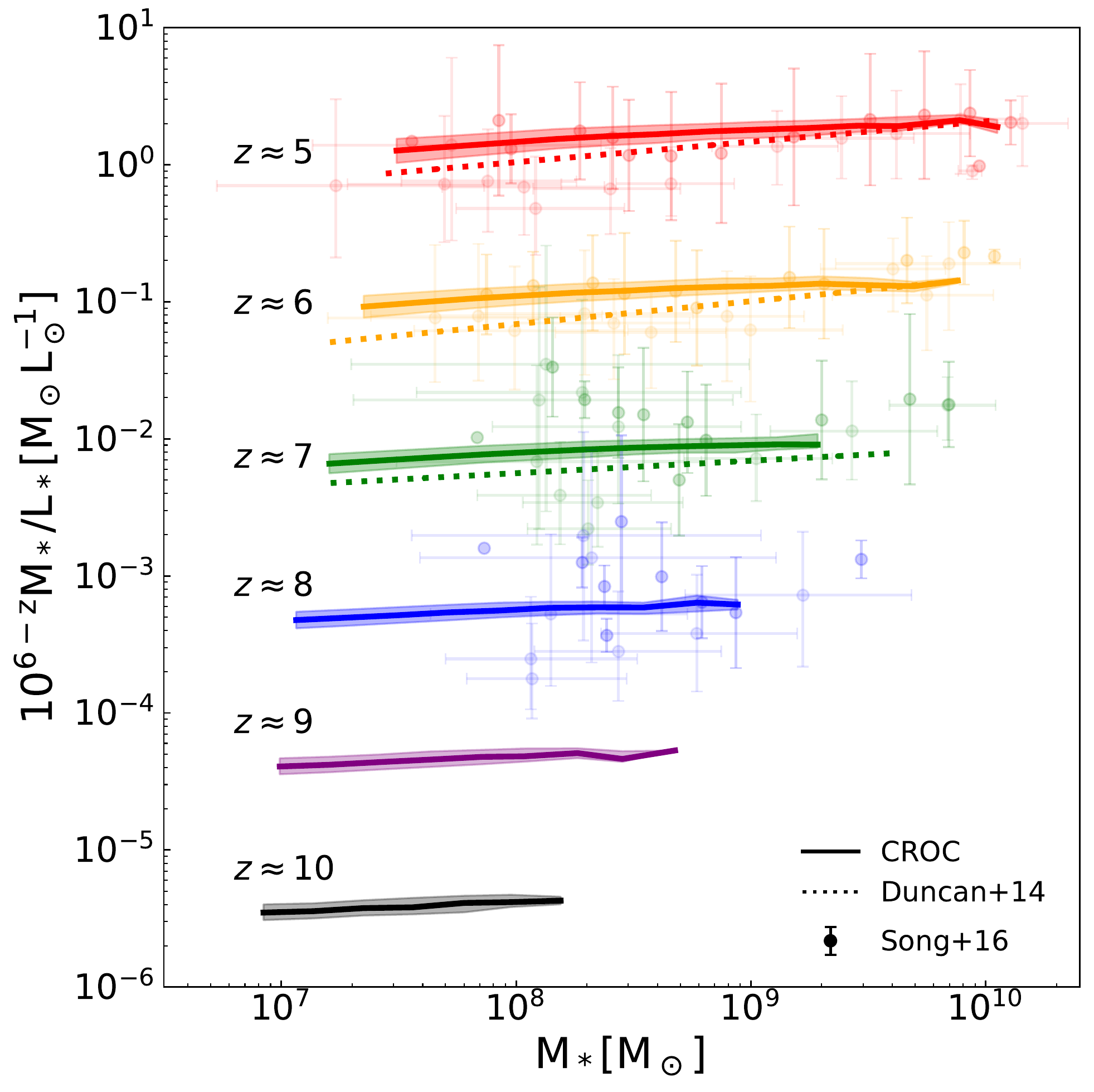}
\caption{Mass-to-light ratio of CROC galaxies as a function of stellar mass. The solid line corresponds to the median CROC ratios in each stellar mass bin, and the transparent bands illustrate the 25th and 75th percentile values in each stellar mass bin. Observational data are from \citet{song16} and \citet{duncan14}.}\label{fig:mass_to_light_ratio}
\end{figure}

In Figure~\ref{fig:mass_to_light_ratio} we show the stellar mass-to-light ratio of galaxies in the CROC simulations along with observational data from \citet{song16} and \citet{duncan14}. The line colors correspond to a redshift range of $5 \leq z\leq10$. We see that, at fixed stellar mass, the mass-to-light ratio increases moderately with time for both the CROC sample and the observations, and the overall agreement is fair. This agreement means that the amount of light produced per unit mass of stars in CROC galaxies (i.e. the shapes of individual star formation histories) is captured faithfully. We emphasize that, while the shapes of individual star formation histories are captured faithfully, this is true at fixed {\it stellar mass}, not at fixed {\it halo mass}. This leads to disagreements with the observed total stellar mass and luminosity, which we discuss in Section~\ref{sec:results:starformationhistory}. On the other hand, the lack of bright galaxies in the simulations is apparent in Figure~\ref{fig:uvlf}, indicating that the overall level of star formation in the most massive galaxies in CROC is lower than that of observed galaxies. In other words, the star formation and stellar feedback model in the simulation works reasonably well in fainter galaxies ($M_{1500}\gtrsim-18$) ($L<L_\star$) but is underproducing stars in brighter, $L>L_\star$  ($M_{1500}\lesssim-18$), galaxies. The same deficiency should therefore manifest itself in the underprediction of SMFs for the most massive galaxies. 

The good agreement for stellar mass-to-light ratios also implies that any potential error due to stellar population synthesis modeling is insignificant: in the simulations we use the FSPS code \citep{fsps}, while observational measurements rely on Bruzual \& Charlot models \citep{bz03,bz11}.

\subsection{Evolution of the SMF}\label{sec:results:stellarmassfunction}
\begin{figure}
    \centering
        \includegraphics[width=0.99\columnwidth]{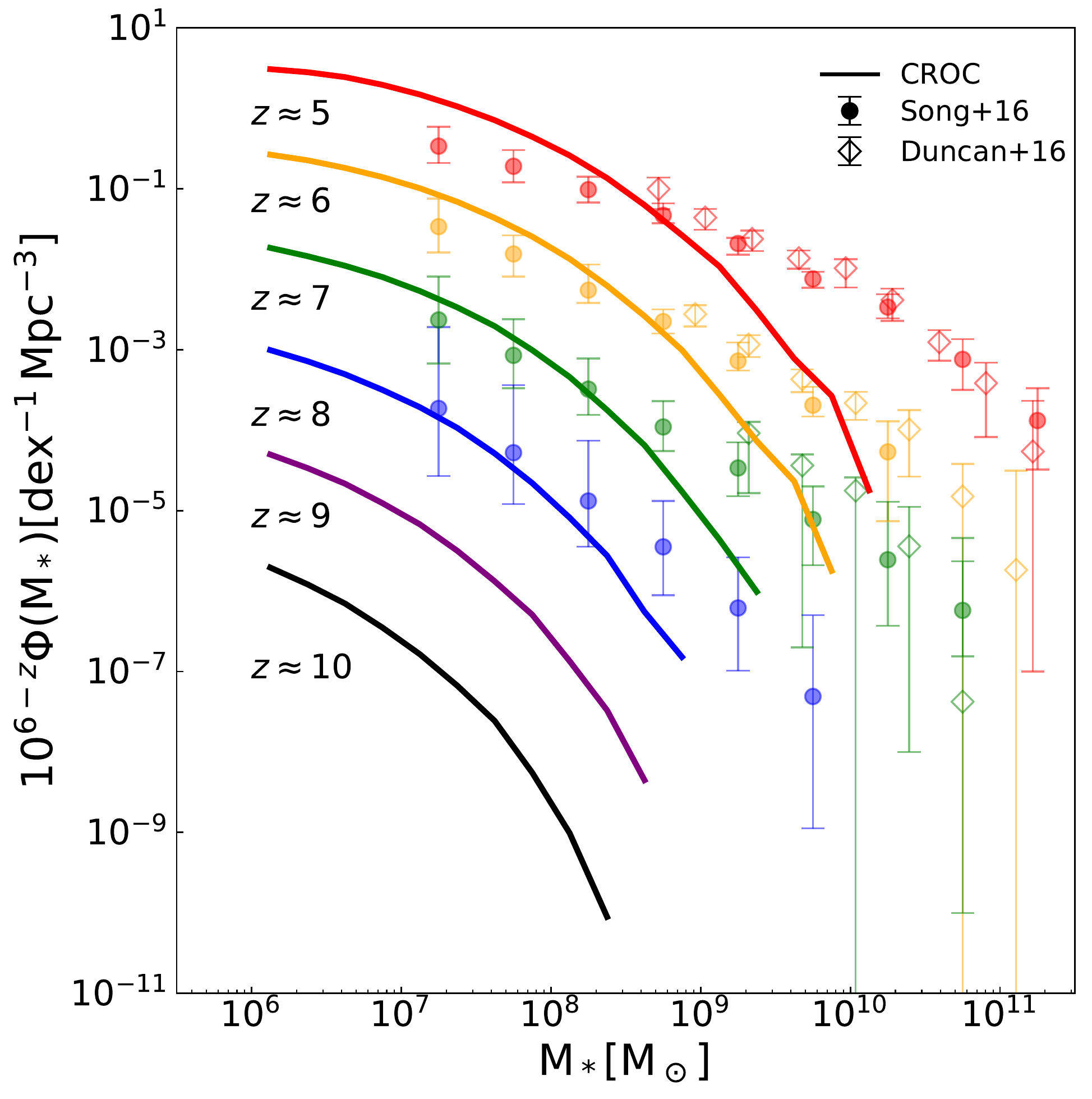} 
\caption{Evolution of stellar mass function of CROC galaxies.  We overplot observational data from \citet{song16} in circles and data from \citet{duncan14} in diamonds.  Relative to observations, CROC has a deficit of high-mass objects at $z<8$.} \label{fig:smf}
\end{figure}

The SMF provides another illustrative comparison between CROC and observations; in the simulations the stellar mass is a primary, directly simulated quantity, while galactic UV luminosity is computed in post-processing. However, this is with the caveat that the observational inference of the SMF is more assumption dependent than the UVLF. Figure~\ref{fig:smf} shows the SMF of CROC galaxies and the observed SMFs from \citet{duncan14} and \citet{song16}.

The SMF exhibits a deficit of massive galaxies, M$_{\star}\gtrsim 3\times10^8\Msun$, which is consistent with the deficit of bright galaxies discussed above.  Since the stellar mass-to-light ratios of massive galaxies in CROC are realistic, the discrepancy we see in the UVLF in Figure~\ref{fig:uvlf} is driven by the inaccuracy of the star formation and stellar feedback model and not, for example, by the errors in computing stellar luminosities. In other words, the galaxies that live in the most massive CROC halos are less massive and less luminous than those actually observed. In order to quantify this further, we consider below the relation between stellar and halo masses in the simulations.

For some applications, this deficiency of the simulations can be ``corrected" in post-processing by re-scaling the simulated stellar masses as
\begin{equation}
  M_\star \rightarrow M_\star\left(1+A_z\log_{10}\left(1+\frac{M_\star}{3\times10^8\Msun}\right)\right),
  \label{eq:rescale}
\end{equation}
where $A_z\approx 6, 10, 30$ at $z=5, 6, 7$. We do not use such re-scaling in this paper, as our goal is to emphasize the regimes where the simulations fail and to identify what needs to be fixed in future simulations, not to simply match all the existing observations. We present this correction here as a reference for the future work.

\subsection{Galaxy Properties and Halo Mass}\label{sec:results:galaxytohalo}

In order to quantify the galaxy-halo connection in CROC, we first examine the SHMR, which quantifies the integrated efficiency of star formation and merger-driven growth.  

\begin{figure}[ht!]
\centering
        \includegraphics[width=0.99\columnwidth]{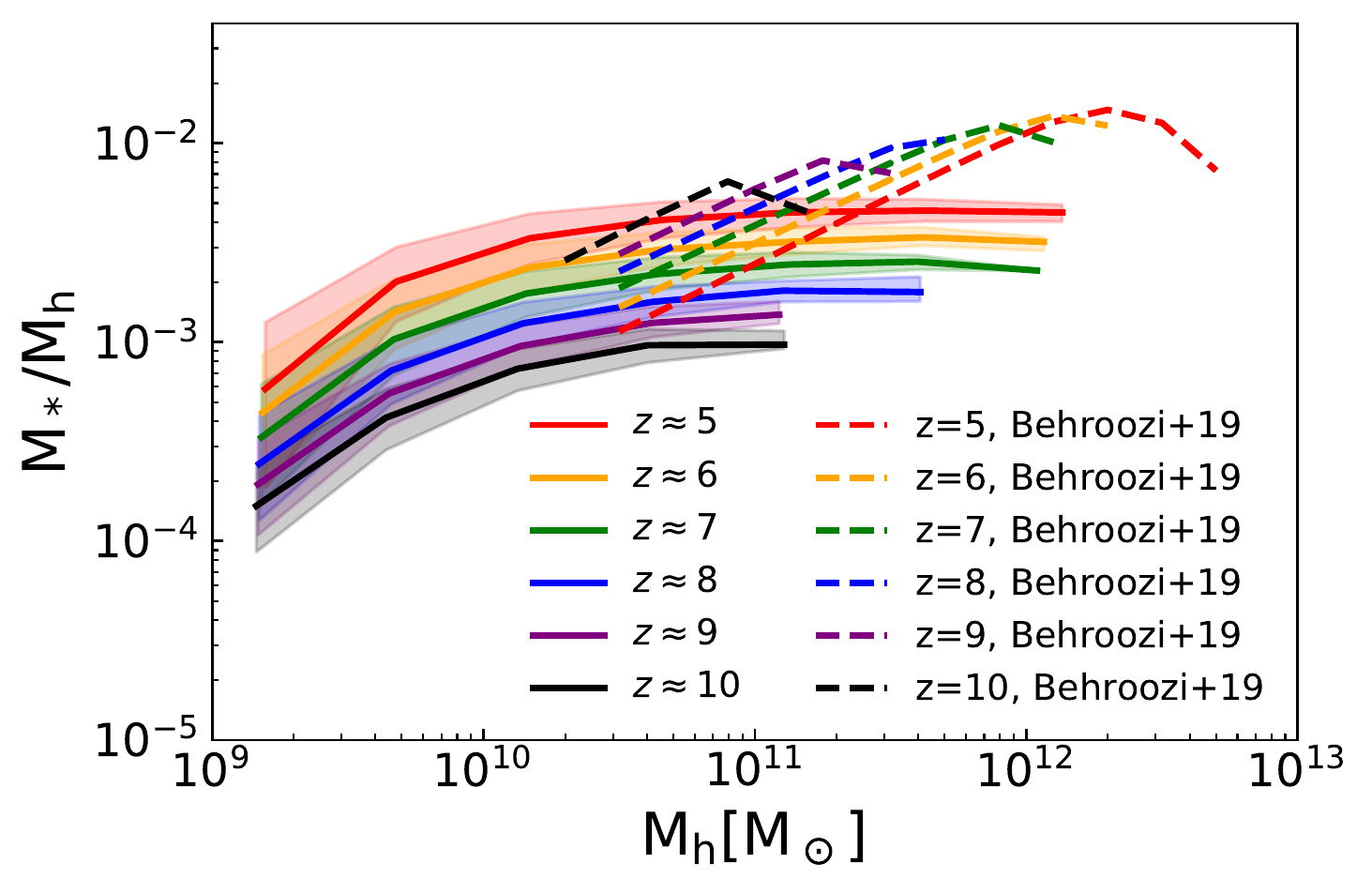}
\caption{Stellar-to-halo-mass ratio (SHMR) as a function of halo mass for CROC simulations (solid lines with bands) from $z=5$ to $z=10$ together with the same relation from \citet{Behroozi19}
obtained from abundance matching.  The SHMR flattens at halo masses above $M_h\gtrsim 10^{10}\Msun$, reflecting the deficiency in massive galaxies apparent in Figs.\ \ref{fig:uvlf} and \ref{fig:smf}. The bands illustrate the 25th and 75th percentile values in each mass bin. }\label{fig:shmr_evolution}
\end{figure}

Figure~\ref{fig:shmr_evolution} shows the SHMR as a function of halo mass between $5 \leq z\leq 10$.  At fixed redshift, the relationship flattens for halos with $M_h\gtrsim 10^{10}\Msun$.  The flattening is consistent with the deficiency in massive galaxies apparent in Figs.\ \ref{fig:uvlf} and \ref{fig:smf}. As host halos grow through hierarchical formation to form the largest objects, galaxies do not seem to be forming more stars.  We explore the evolution of star formation separately in subsequent sections.

Separate from the flattening of the SHMR is the redshift evolution, which is somewhat of an open question given discrepancies between models and observations. At fixed halo mass, the SHMR of CROC galaxies decreases with increasing redshift, indicating the star formation efficiency in galaxies is lower in the past. This is consistent with the results from \citet{tacchella_etal18}, where they assume a time delay for star formation that has a larger impact at higher redshifts. 

This evolutionary trend, however, is contrary to the observational findings of \citet{finkelstein_etalb15}, \citet{harikane_etal16}, and \citet{stefanon17}. Both \citet{finkelstein_etalb15} and \citet{harikane_etal16} found evidence that the SHMR {\it increases} from $z=4$ to $z=7$.  On the other hand, \citet{stefanon17} found no evidence of evolution between $4\leq z\leq7$, consistent with model predictions from the BLUETIDES simulation \citep{Wilkins2017,bhowmick_etal18} and from the FIRE-2 simulations \citep{ma18}.  

The observational methods for constraining the SHMR vary; \citet{harikane_etal16} combined a halo occupation distribution model for angular correlation functions to constrain the SHMR with measurements of stellar mass, and \citet{stefanon17} used an abundance-matching method.  Additionally, the uncertainties associated with the observed SHMR at high redshifts have the same sources as those in the SMF, where stellar masses are derived from empirical relations with large intrinsic scatter.  These factors could explain the disparity across observations in constraining the SHMR, and much improved constraints can be expected from the forthcoming JWST observations, whose increased sensitivity and observational area will yield data with tighter errorbars at lower luminosities.

The comparison with observational constraints from \citet{finkelstein_etalb15} and \citet{harikane_etal16} needs to be done with care as the observations only sample halo masses in excess of $10^{11}\Msun$ ($M_*>3\times10^8\Msun$), where CROC simulations fail. As an illustration, however, we can apply the correction from Equation (\ref{eq:rescale}) to extend simulation results to higher stellar masses. Surprisingly, we find that we are unable to match \citet{harikane_etal16} with corrections (\ref{eq:rescale}) that are chosen to reproduce the \citet{duncan14} measurement. In fact, factors $A_z$ in Equation (\ref{eq:rescale}) would have to be three times smaller to match \citet{harikane_etal16}. In other words, we are unable to populate CROC dark matter halos with stars with \emph{any} $M_*(M_h)$ in such a way as to match simultaneously \citet{harikane_etal16} and \citet{duncan14}.

For comparison, in Figure~\ref{fig:shmr_evolution} we show model predictions from abundance matching \citep{Behroozi19} with dashed lines. It is interesting to note that for halo masses below $M_h\lesssim5\times10^{11}\Msun$, the \citet{Behroozi19} model predicts that the SHMR {\it decreases} with from $z=5$ to $z=8$, but above this mass range, the redshift trend reverses. Despite the fact that both approaches reproduce the observed galaxy UVLF and SMF in a given mass range (for CROC only for $M_\star<10^8\Msun$), CROC and \citet{Behroozi19} display opposite evolution trends in the SHMR. This implies that the observational errors in the UVLF and SMF are still large enough to allow \emph{opposite} trends in the SHMR to be consistent with the data.

\subsection{Star formation over cosmic time}\label{sec:results:starformationhistory}

\begin{figure}[ht!]
\centering
        \includegraphics[width=0.99\columnwidth]{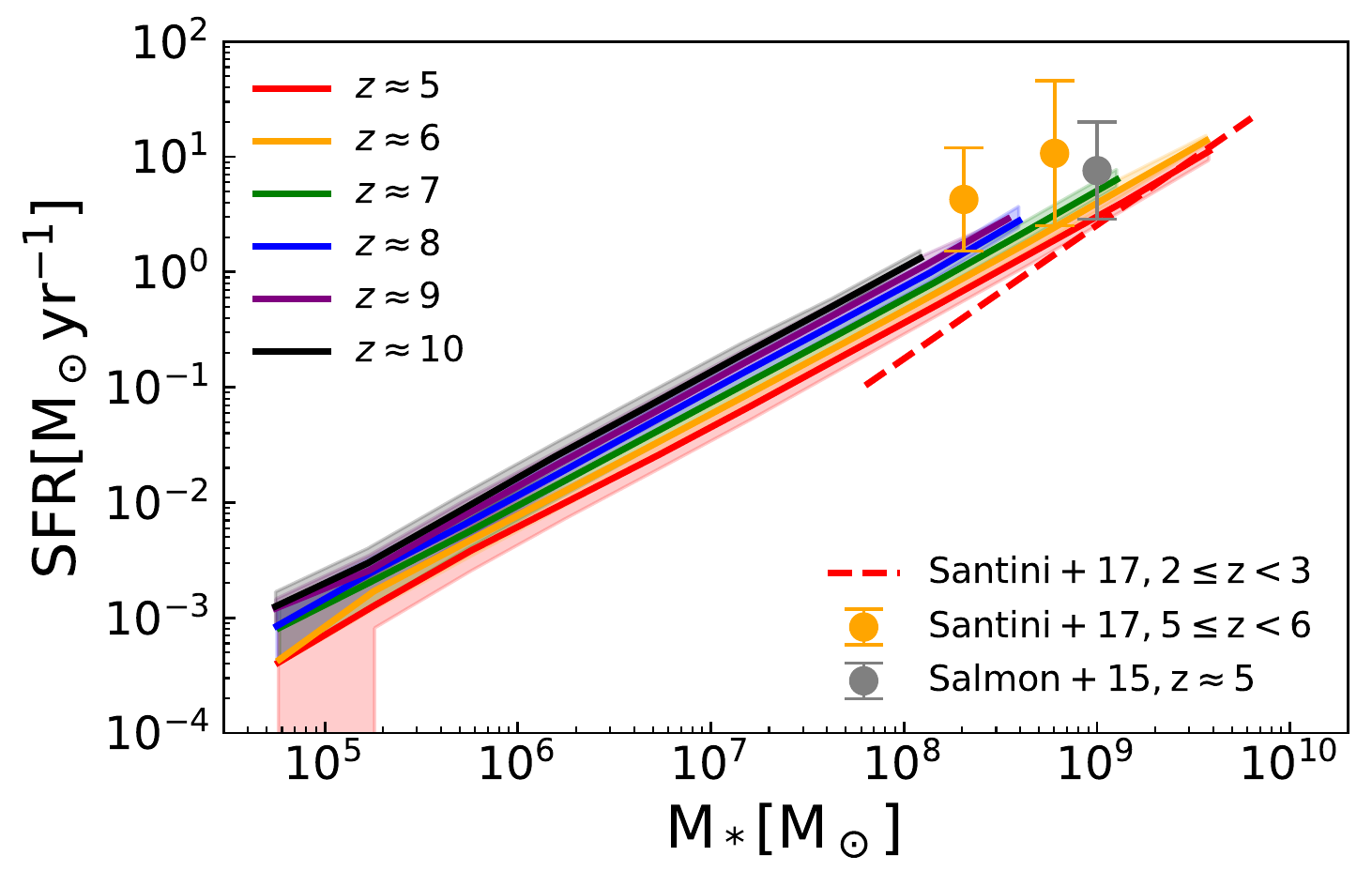}
\caption{Star formation rate as a function of stellar mass from $z=5$ to $z=10$. The bands illustrate the 25th and 75th percentile values in each mass bin.  We also show the relation from observations in \citet{santini17}, where the slope is similar and a similar evolutionary trend exists.  The $z=5$ observed slope matches those in CROC, but the observed normalization is higher. Also, note that the observed slope steepens with time.     \label{fig:sfr_evolution}}
\end{figure}

The deficiency at high stellar masses can be better understood by looking at how the star formation rate at fixed stellar mass varies with redshift, otherwise known as the evolution of the ``main sequence of galaxy formation." The solid lines in Figure~\ref{fig:sfr_evolution} show that the star formation rate at a fixed stellar mass {\it decreases} with redshift, consistent with the trends generally found in observational studies \citep[see][]{santini17}. At a given redshift, the slope of the star formation rate-$M_\star$ relation is sub-linear and is slightly shallower than the slope of the observed relation at $2\leq z<3$. 

The star formation rate in galaxies has been relatively well measured by several teams over the last few years, such as \citet{salim07,catalantorrecilla15,pannella15}, and \citet{zahid17} for low-redshift galaxies and \citet{whitaker12} for galaxies up to $z=2.5$. At even higher redshifts the measurements are limited by a number of systematic effects, such as those present in gravitationally lensed systems that magnify and enable measurements of high-redshift objects. Therefore, the robustness of the relationship between star formation rate and stellar mass at high redshifts is not yet well established.  In general our galaxies form stars at rates consistent with, but perhaps somewhat lower than, observations. Note, however, that any difference between CROC and observed galaxy star formation rates is much smaller than the correction (Eq.\ \ref{eq:rescale}) required to bring CROC SMFs in agreement with observational data.

\begin{figure}
    \centering
        \includegraphics[width=0.99\columnwidth]{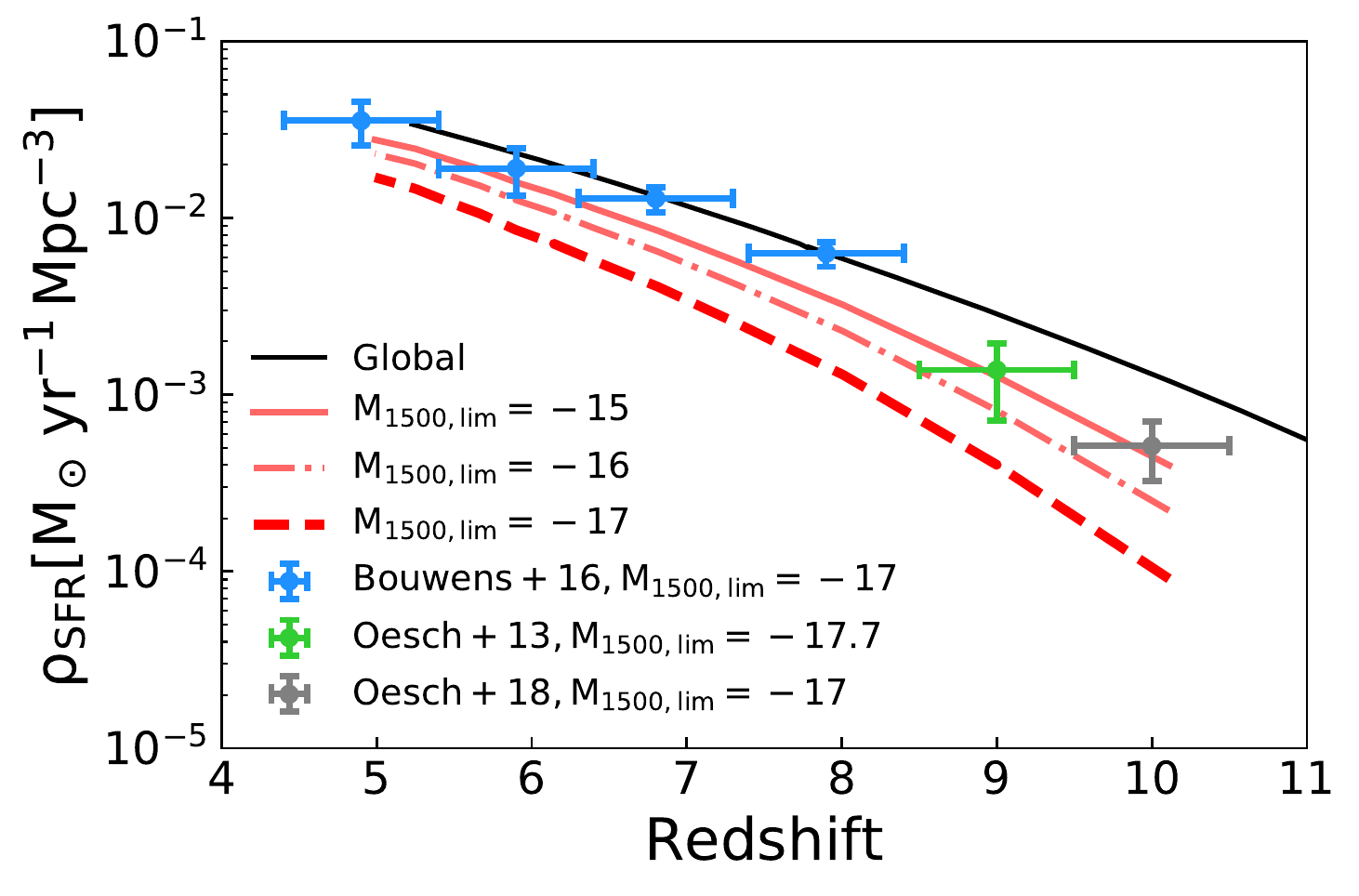} 
\caption{Evolution of star formation rate density (SFRD) in CROC compared with data from \citet{bouwens16} in blue errorbars, \citet{oesch13} in the green errorbar, and \citet{oesch18} in the gray errorbar. The solid black line is derived from the global star formation rate, while the red solid, dotted-dashed, and dashed lines show the results obtained by having different \Muv \ cuts.}\label{fig:sfr_density}
\end{figure}

In Figure~\ref{fig:sfr_density} we show the evolution of the global star formation rate density (SFRD). First, we see that the SFRD decreases as redshift increases.  Next, as we include fainter galaxies into our calculation of SFRD, it increases as expected. In order to compare with observations, we integrate the UVLFs above the magnitude cutoff of \Muv \ $\approx-17$. Compared to simulations, the observational results give higher SFRD values at all redshifts. This is another illustration that that CROC is underproducing stars in massive galaxies at all cosmic times.

\begin{figure}
    \centering
        \includegraphics[width=0.99\columnwidth]{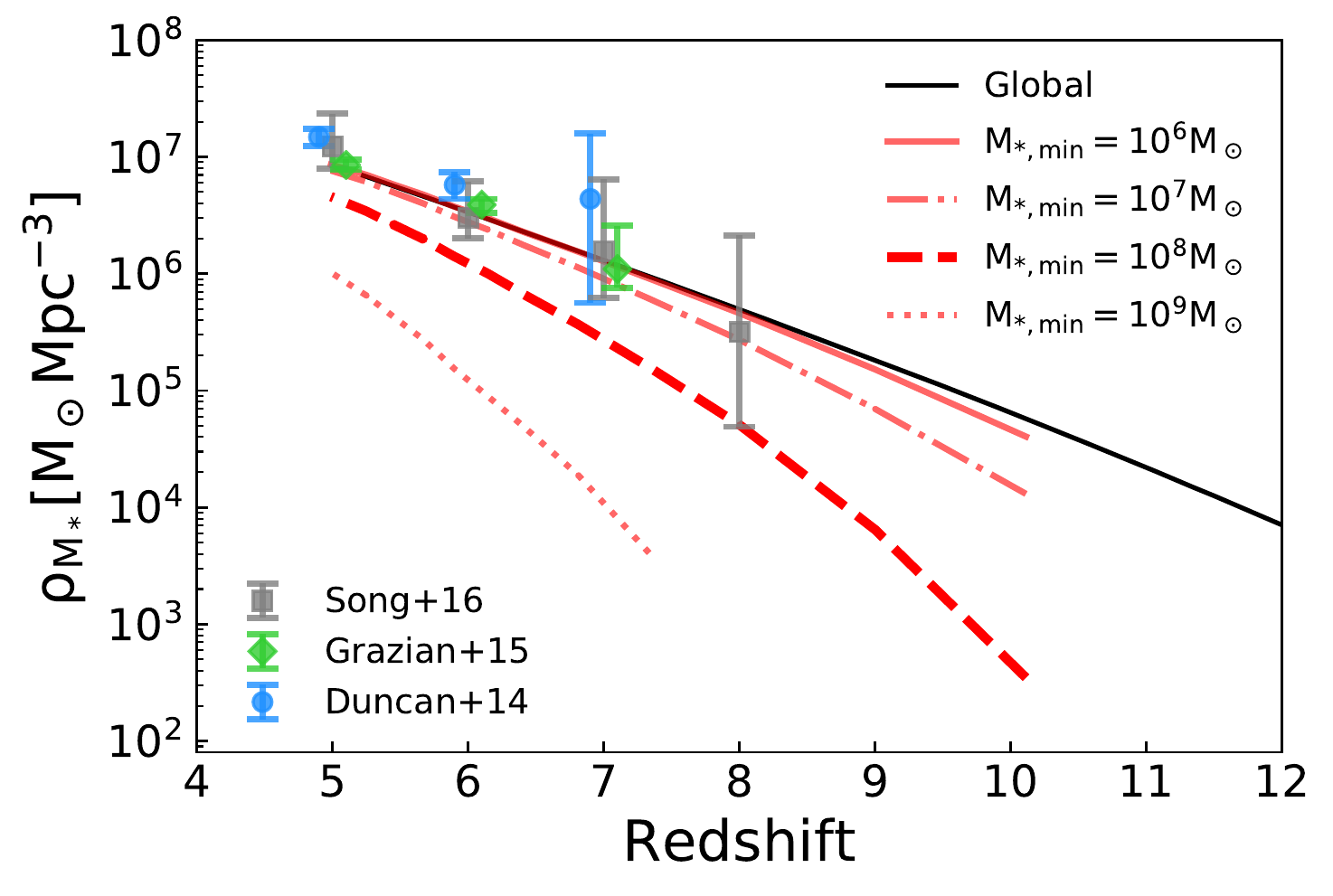} 
\caption{Evolution of stellar mass density in CROC with data from \citet{song16} in gray errorbars, \citet{grazian15} in green errorbars, and \citet{duncan14} in blue errorbars. These observational results are derived by integrating their best-fit stellar mass functions above $M_{\star} > 10^{8} \Msun$. We show the CROC results obtained by including halos with different lower stellar mass limits, including the $10^{8}\Msun$ limit as the dotted–dashed line.}\label{fig:stellar_mass_density}
\end{figure}

The corresponding evolution of the global stellar mass density is shown in Figure~\ref{fig:stellar_mass_density}. The global stellar mass density is harder to measure than the global star formation rate.  The observational uncertainties are therefore larger. Still, the underprediction of stellar mass in CROC simulations in galaxies above $10^8\Msun$ (thick dashed red line) is very apparent. 

\subsection{Galaxy Bias}\label{sec:results:bias}

The galaxy bias measurement provides an additional constraint on the galaxy-halo connection, measuring the excess clustering of galaxies compared with the clustering of matter.  The bias can be calculated in terms of the two-point correlation function, $\langle b^2_{\mathrm{gal}}\rangle={\xi_{gg}}/{\xi_{mm}}$, where $\xi$ is the two-point correlation function of either galaxies (gg) or matter (mm).  When compared to observed galaxies, we can constrain whether or not our modeled galaxies have similar spatial clustering as the observed galaxies. Given consistent cosmologies, we can assume that our simulated halo bias matches that in the real universe, $\langle b^2_\mathrm{halo}\rangle={\xi_{hh}}/{\xi_{mm}}$.  We can then use the galaxy bias comparison to determine whether or not CROC galaxies live in the right dark matter halos.

In order to calculate the galaxy bias across mass bins including the bins of the most massive galaxies, whose number counts are small, we can use the empirical form of the bias dependence on the halo mass, $b(M_h)$, and compute the average galaxy bias for a given sample as an integral over the halo mass function, instead of actually computing the ratio of two power spectra. We note that this method assumes that galaxies in a luminosity bin constitute a uniform sub-sample of halos in a corresponding mass bin, which is consistent with Figure~\ref{fig:muv_mh}.
For the explicit calculation, we use the \citet{tinker_10} halo bias model, $b(M_h)$, from the cosmology package Colossus \citep{colossus}. The simulated halo mass function of CROC well matches the halo mass function from \citet{tinker_10}, the corresponding bias model is therefore the appropriate model to use for CROC halos. As an example, for a halo sample with halo masses between $M_\mathrm{min}$ and $M_\mathrm{max}$ the average bias is 
\begin{equation}
\langle b\rangle =
\frac{\int^{M_{\mathrm{max}}}_{M_{\mathrm{min}}}b(M_h)\frac{dn}{dM_h}dM_h}{\int_{M_{\mathrm{min}}}^{M_{\mathrm{max}}}\frac{dn}{dM_h}dM_h}.
\label{eq:biash}
\end{equation}
For a galaxy sample containing $N$ galaxies (with an arbitrary sample selection criterion) Equation (\ref{eq:biash}) reduces to the average of the bias factors of individual galaxies:
\begin{equation}
\langle b\rangle = \frac{1}{N} \sum_{i=1}^N b(M_{h,i}),
\label{eq:bias}
\end{equation}
where $M_{h,i}$ is the known halo mass for galaxy $i$ from the sample.

\begin{figure}[th]
    \centering
        \includegraphics[width=0.99\columnwidth]{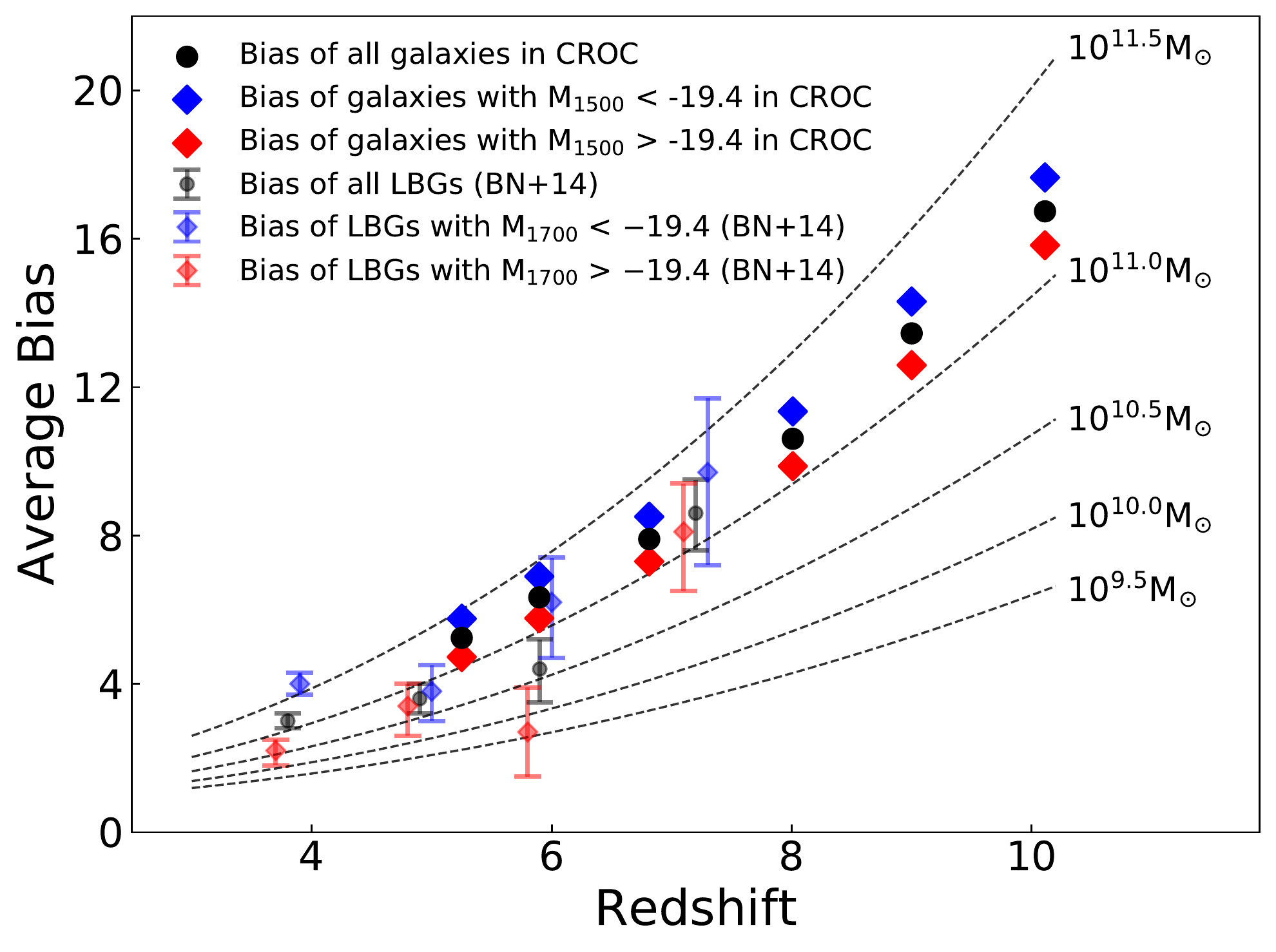} 
\caption{Galaxy bias in the CROC simulations compared with measurements from \citet{baronenugent_etal14} and dark matter halo bias from \citet{tinker_10}. We illustrate the bias in two magnitude bins, brighter (blue) and fainter (red) than $M_{1500}=-19.4$. As expected, the brighter galaxies exhibit a higher average bias in each redshift snapshot.  The CROC measurements are consistent with observations.  However, the faint bin of CROC galaxies exhibit a somewhat higher bias than the observed value at z$\leq$6 since we are using the brightest CROC galaxies in the $M_{1500}>-19.4$ bin}.\label{fig:bias}
\end{figure}

Figure~\ref{fig:bias} shows the so-computed average {\it galaxy} bias from all six boxes in the CROC suite.
The blue and red squares respectively show the average galaxy bias for bright ($M_{1500}<-19.4$) and faint ($M_{1500}>-19.4$) galaxies in CROC as a function of redshift. The black squares show the average galaxy bias for all galaxies in the simulation as a function of redshift.  For comparison, the points with errorbars show observational inferences of the bias from \citet{baronenugent_etal14} with the same UV magnitude selection. There is good agreement between the simulations and the data, especially at $z \sim 7$, which indicates two things. First, CROC galaxies that have lower luminosities than the luminosity cutoff of $M_{1500}=-19.4$ match the observed luminosity and SMFs well. Second, the clustering of dark matter halos is captured properly in the simulations.

Given the assumption that the halo bias in simulations reflects that of the real universe, and the fact that the faint (low-mass) end of the CROC luminosity (stellar mass) function matches the observed function, we can first say that the low-luminosity sample of CROC galaxies below $M_{1500}=-19.4$ live in the correct halos and well reproduce the observed behavior.

The minimum stellar mass of galaxies brighter that $M_{1500}=-19.4$ corresponds to $M_\star\approx 3\times10^8\Msun$.  These brighter galaxies above the luminosity cutoff also seem to well reproduce the observed galaxy bias, illustrating our second point, that the bias factor from Equation~\ref{eq:bias} has been modeled correctly.  However, this population of galaxies have UVLFs that are discrepant with the observed UVLF.  In other words, the high-mass halos have galaxies in the correct luminosity bin brighter than $M_{1500}=-19.4$, but there are not enough stars in these galaxies to match the UVLF at the bright end.

Note that the galaxy bias of the fainter galaxy bin may appear to be on the high side.  However, we use the same number of CROC galaxies whose luminosities are below the cutoff as the number of CROC galaxies that lie above the cutoff, selecting the brightest galaxies when rank ordered by luminosity. The resulting faint galaxy sample is comprised of galaxies whose (1) total masses are $M_{\rm tot} \gtrsim 10^{11}\Msun$ and (2) magnitudes are $M_{1500}<-19.0$ at $z=5$, gradually decreasing to $M_{1500}<-19.2$ by $z=10$. The galaxy bias in this bin is therefore larger than the average bias of all galaxies with $M_{1500}>-19.4$ bin.

\section{Summary and Discussion}
In this paper, we examine the galaxy-halo connection of halos in the mass range $\rm M_{halo} \approx 10^9-10^{12}\Msun$ at $5\leq z\leq10$ in CROC simulations. CROC simulations include a wide range of physical processes necessary to model cosmic reionization, including gravity, gas dynamics, fully coupled radiative transfer, radiative cooling, star formation, and stellar feedback. We look at the UVLF, the SMF, the SHMR, star formation histories, and the galaxy bias in CROC.

Our main findings include the following.
\begin{itemize}
\item CROC galaxies match the UVLF and the SMF at the faint and low-mass end respectively.
\item  Our simulations underproduce stellar masses of the most massive halos in the CROC simulations ($\approx10^{11}\Msun$). Physical processes that are not modeled in CROC, such as AGN feedback and cosmic ray feedback, can reduce star formation rates in massive halos, further exacerbating the discrepancy. Hence the most likely reason for the discrepancy is that the CROC model for star formation and feedback mis-models star formation in high-mass halos. The addition of AGNs is likely to exacerbate the over-suppression of star formation in massive galaxies, while not significantly affecting fainter galaxies, since it is generally believed that AGNs primarily impact more massive objects. Hence, including the AGN feedback is unlikely to help with improving the overall shape of the luminosity function. This indicates that the simulations altogether require a different model for stellar feedback that does not over-suppress star formation in the high-mass objects. Another, admittedly much less likely, possibility is that the discrepancy between CROC and observations in the high-mass end is due to a bias in observational results from, for example, AGN contributions to galactic UV luminosities \citep{Adams2020}. 
\item The SHMR in CROC decreases with redshift, consistent with some other models, but inconsistent with abundance-matching predictions from \citet{Behroozi19} and some of the observed trends. In fact, we are unable to populate CROC dark matter halos with stars with \emph{any} function form of $M_*(M_h)$ in such a way as to match simultaneously \citet{harikane_etal16} and \citet{duncan14}. Since all models and observational constraints match the observed luminosity functions and use the same cosmology, the lack of consistency even in the sign of the redshift evolution implies that the observed luminosity functions are not accurate enough to constrain the SHMR at $z>5$.
\end{itemize}

These results will help focus efforts on improving the accuracy and physical fidelity of future simulations of cosmic reionization.

\acknowledgments
This paper was substantially improved with the comments from an anonymous referee; the authors also thank the referee for catching a serious error in the original manuscript. The authors are grateful to Mimi Song and Kenneth Duncan, who enabled the comparison of the CROC simulations with observational data. The authors thank Huanqing Chen and Hsiao-Wen Chen for helpful discussions that guided the analysis and interpretation of results, and Andrey Kravtsov and Phil Mansfield for extensive and comprehensive comments that allowed us to significantly improve the original manuscript. H.Z. acknowledges support from the Kavli Institute for Cosmological Physics and the Jeff Metcalf Fellowship Program. C.A. acknowledges support from both the Enrico Fermi Institute the Kavli Institute for Cosmological Physics at the University of Chicago,  and both the Leinweber Center for Theoretical Physics and the LSA Collegiate Fellowship at the University of Michigan. Fermilab is operated by Fermi Research Alliance, LLC, under Contract No. DE-AC02-07CH11359 with the United States Department of Energy. This work was partly supported by a NASA ATP grant NNX17AK65G, and used resources of the Argonne Leadership Computing Facility, which is a DOE Office of Science User Facility supported under Contract DE-AC02-06CH11357. An award of computer time was provided by the Innovative and Novel Computational Impact on Theory and Experiment (INCITE) program. This research is also part of the Blue Waters sustained-petascale computing project, which is supported by the National Science Foundation (awards OCI-0725070 and ACI-1238993) and the state of Illinois. Blue Waters is a joint effort of the University of Illinois at Urbana-Champaign and its National Center for Supercomputing Applications.

\bibliographystyle{apj}
\bibliography{main}

\end{CJK*}
\end{document}